# Domain Decomposition Based High Performance Parallel Computing

**Mandhapati P. Raju[1] and Siddhartha Khaitan[2]**

**[1] Mechanical Engineering, Case Western Reserve University,
Cleveland, Ohio 44106**

**[2] Electrical and Computer Engineering, Iowa State University,
Ames, IA 50011**

## Abstract

The study deals with the parallelization of finite element based Navier-Stokes codes using domain decomposition and state-of-art sparse direct solvers. There has been significant improvement in the performance of sparse direct solvers. Parallel sparse direct solvers are not found to exhibit good scalability. Hence, the parallelization of sparse direct solvers is done using domain decomposition techniques. A highly efficient sparse direct solver PARDISO is used in this study. The scalability of both Newton and modified Newton algorithms are tested.

## 1. Introduction

Three dimensional finite element problems pose severe limitations in terms of computational time and memory requirement. The choice of a direct solver or an iterative solver for large problems is not trivial. Iterative solvers are considered to be highly memory efficient, especially for larger problems. However, it is observed that the choice of an iterative solver is highly problem specific. For problems involving highly ill conditioned matrices, iterative solvers do not perform well. Direct solvers on the other hand are considered to be very robust and problem independent. However, they are seriously limited by large memory requirements. Memory issue for direct solvers is being addressed through several ways. The advent of multifrontal solvers [1] coupled with highly efficient ordering techniques has greatly increased the computational efficiency and reduced the memory requirement for direct solvers. The superior performance of Multifrontal solvers has been successfully demonstrated both in the context of finite volume problems [2-4], finite element problems [5-6] and in power system simulations

[7-9]. For three dimensional problems the memory requirement is still a limitation for larger problems. To circumvent this problem, a 64-bit machine with a larger RAM can be used for computation, an out-of-core sparse direct solvers can be used which have the capability of storing the factors on the disk during factorization or the direct solvers can be used in a distributed computing environment. This paper specifically deals with the application of sparse direct solvers in a distributed computing environment.

Recently a parallel direct solver, MUltifrontal Massively Parallel Solver (MUMPS) [10-12] in a distributed environment [6] has been studied for both two dimensional and three dimensional problems. It has been reported that by using MUMPS solver, larger problems can be solved in a parallel environment as the memory is distributed amongst the different processors. However it has also been reported that the scalability of MUMPS solver [6] is not very high. In order to obtain good scalability, instead of using a parallel sparse direct solver, the problem itself is split amongst the different processors using domain decomposition [13] and a highly efficient sparse direct solver is used in each of subdomain problems. An efficient sparse direct solver PARDISO [14-16] is being used in this study.

Laminar flow through a rectangular channel has been chosen as a benchmark problem for studying the scalability of domain decomposition techniques in a distributed environment. The system of non-linear equations obtained from the discretization of Navier-Stokes equations is solved using Newton's method. The system of linear equations obtained during each Newton iteration is solved in parallel using additive Schwarz domain decomposition algorithm [13]. The grid is divided into many overlapping subdomains (equal to the number of processors being used) and each processor solves its local subdomain problem using an efficient sparse direct solver. Figure 1 shows a typical overlapping subdomain for additive Schwarz algorithm. In additive Schwarz algorithm, the values at the interfaces are updated only after a complete cycle of subdomain solution is computed. In using direct solvers, factorization of the left hand side matrix is the most time consuming step. To avoid repeated factorization in the subdomains, advantage is taken of the resolution facility of the direct solvers (i.e. the previously computed $LU$ factors are reused in the solve phase). A modified version of the additive Schwarz algorithm is also proposed which is computationally very efficient.

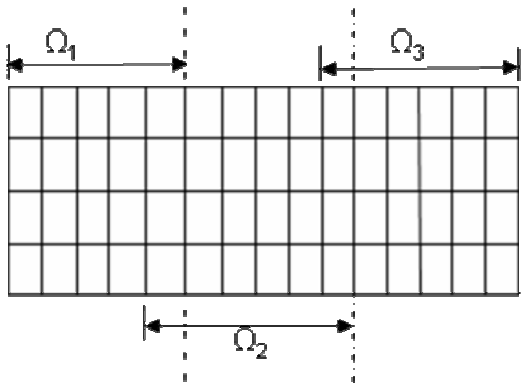

Figure 1: Typical overlapping subdomain division for additive Schwarz algorithm

## 2. Mathematical Formulation

The governing equations for laminar flow through a three-dimensional rectangular duct [6] are presented below in the non-dimensional form. In three-dimensional calculations, instead of the primitive $u,v,w$ formulation, penalty approach is used to reduce the memory requirements.

$$\frac{\partial u}{\partial x} + \frac{\partial v}{\partial y} + \frac{\partial w}{\partial z} = 0, \tag{1}$$

$$\frac{\partial}{\partial x}\left(u^2\right) + \frac{\partial}{\partial y}(uv) + \frac{\partial}{\partial z}(uw) = \lambda \frac{\partial}{\partial x}\left(\frac{\partial u}{\partial x} + \frac{\partial v}{\partial y} + \frac{\partial w}{\partial z}\right) + \frac{\partial}{\partial x}\left(\frac{2}{\mathrm{Re}}\frac{\partial u}{\partial x}\right)$$
$$+ \frac{\partial}{\partial y}\left(\frac{1}{\mathrm{Re}}\left(\frac{\partial u}{\partial y} + \frac{\partial v}{\partial x}\right)\right) + \frac{\partial}{\partial z}\left(\frac{1}{\mathrm{Re}}\left(\frac{\partial u}{\partial z} + \frac{\partial w}{\partial x}\right)\right), \tag{2}$$

$$\frac{\partial}{\partial x}(uv) + \frac{\partial}{\partial y}\left(v^2\right) + \frac{\partial}{\partial z}(vw) = \lambda \frac{\partial}{\partial y}\left(\frac{\partial u}{\partial x} + \frac{\partial v}{\partial y} + \frac{\partial w}{\partial z}\right) + \frac{\partial}{\partial x}\left(\frac{1}{\mathrm{Re}}\left(\frac{\partial u}{\partial y} + \frac{\partial v}{\partial x}\right)\right)$$
$$+ \frac{\partial}{\partial y}\left(\frac{2}{\mathrm{Re}}\frac{\partial v}{\partial y}\right) + \frac{\partial}{\partial z}\left(\frac{1}{\mathrm{Re}}\left(\frac{\partial v}{\partial z} + \frac{\partial w}{\partial y}\right)\right), \tag{3}$$

and

$$\frac{\partial}{\partial x}(uw) + \frac{\partial}{\partial y}(vw) + \frac{\partial}{\partial z}\left(w^2\right) = \lambda \frac{\partial}{\partial z}\left(\frac{\partial u}{\partial x} + \frac{\partial v}{\partial y} + \frac{\partial w}{\partial z}\right) + \frac{\partial}{\partial x}\left(\frac{1}{\mathrm{Re}}\left(\frac{\partial u}{\partial z} + \frac{\partial w}{\partial x}\right)\right)$$
$$+ \frac{\partial}{\partial y}\left(\frac{1}{\mathrm{Re}}\left(\frac{\partial v}{\partial z} + \frac{\partial w}{\partial y}\right)\right) + \frac{\partial}{\partial z}\left(\frac{2}{\mathrm{Re}}\frac{\partial w}{\partial z}\right). \tag{4}$$

where $u, v, w$ are the $x$, $y$ and $z$ components of velocity, the bulk flow Reynolds number, $Re = \rho U_0 D / \mu$, $U_0$ being the inlet velocity, $\rho$ the density, $L$ the channel length, $\mu$ is the dynamic viscosity and $\lambda$ is the penalty parameter. Velocities are non-dimensionalised with respect to $U_0$.
The boundary conditions are prescribed as follows:
(1) Along the channel inlet:

$$u = 1; \; v = 0; w = 0. \tag{5}$$

(2) Along the channel exit:

$$\frac{\partial u}{\partial x} = 0; \; \frac{\partial v}{\partial x} = 0; \; \frac{\partial w}{\partial x} = 0. \tag{6}$$

(3) Along the walls:

$$u = 0; \; v = 0; \; w = 0. \tag{7}$$

## 3. Newton's Algorithm

The above set of differential equations (Eq. 1-4) is discretized using Galerkin finite element formulation (GFEM). This results in a set of non-linear equations which is solved using Newton's iterative algorithm.
Let $\underline{X}^{(k)}$ be the available vector of field unknowns for the $(k+1)^{\text{th}}$ iteration. Then the update is obtained as

$$\underline{X}^{(k+1)} = \underline{X}^{(k)} + \alpha \, \delta \underline{X}^{(k)}, \tag{8}$$

where $\alpha$ is an under-relaxation factor, and $\delta \underline{X}^{(k)}$ is the correction vector obtained by solving the linearized system

$$[J]^{(k)}\left\{\delta \underline{X}^{(k)}\right\} = -\left\{F\right\}^{(k)}. \tag{9}$$

Here, $[J]$ is the Jacobian matrix,

$$[J]^{(k)} = \frac{\partial F^{(k)}}{\partial \underline{X}^{(k)}}. \tag{10}$$

and $\{F\}^{(k)}$ is the residual vector. Newton's iteration is continued till the infinity norm of the correction vector $\delta X^{(k)}$ converges to a prescribed tolerance of $10^{-6}$.

# 4. Domain Decomposition

The solution of linear equations (Eq. 9) resulting during each Newton step is the most time consuming time step. It is both computationally expensive and memory consuming. To handle the computational burden, Eq. (9) is solved in a parallel distributed environment using additive Schwarz domain decomposition technique. The domain is split into approximately equal size (similar to Figure 1) subdomains (equal to the number of processors) and distributed amongst the different processors. Each processor solves it own subdomain problem. The partitioning of the domain into subdomains is done by using METIS [17] partitioning routine. The additive Schwarz procedure [13] is similar to the block-Jacobi iteration and consists of updating all the block components from the same residual. The basic additive Schwarz iteration is given below in brief.

$$
\begin{aligned}
&1.\ A = J^{(k)};\quad b = F^{(k)};\quad p = \delta X^{(k)}\\
&2.\ \text{For } i = 1,\dots,s\\
&3.\ \delta_i = R_i^T A_i^{-1} R_i \left(b - Ap\right)\\
&4.\ \text{Enddo}\\
&5.\ p_{new} = p + \alpha \sum_{i=1}^{s} \delta_i
\end{aligned}
\tag{11}
$$

Here $R_i$ is the restriction operator, which restricts a given vector of global domain $\Omega$ to a local subdomain $\Omega_i$. $R_i^T$ is a prolongation operator which takes a variable from local subdomain $\Omega_i$ to global domain $\Omega$. Each instance of the loop redefines different components of the new approximation and there is no data dependency between the subproblems involved in the loop. Each subproblem is solved on a different processor. Instead of solving the whole problem (eq. 9) on a single processor, the problem is divided into subdomain problems (step 3 of eq. 11) which each processor can solve independently. Each subdomain problem is solved using an efficient sparse direct solver, PARDISO. $\alpha$ is the under-relaxation factor. The value of under-relaxation factor is chosen as 1 for updating all variables at the interior nodes of the subdomain and 0.6 for updating the variables at the subdomain interface nodes. The system is solved using two iteration loops. The first in the outer Newton iterative loop and the second is the inner iterative loop for solving the system of linear equations (Eq. 9) using additive Schwarz algorithm (Eq. 11). This algorithm is being referred in this paper as Newton based additive Schwarz algorithm (NAS).

To improve the computational efficiency of the NAS algorithm, a modified Newton based additive Schwarz algorithm (MNAS) is proposed. The algorithm is presented below.

$$
\begin{aligned}
&1.\ A = J^{(k)};\quad A^0 = J^{(0)};\quad b = F^{(k)};\quad p = \delta X^{(k)}\\
&2.\ \text{For } i = 1,\dots,s\\
&3.\ \delta_i = R_i^T \left(A^0\right)_i^{-1} R_i \left(b - Ap\right)\\
&4.\ \text{Enddo}\\
&5.\ p_{new} = p + \alpha \sum_{i=1}^{s} \delta_i
\end{aligned}
\tag{12}
$$

In essence, it solves the same system of equations as the NAS algorithm, since $A$ and $b$ are the same. Consequently the quadratic Newton convergence is retained. The difference is that the step 3 of additive Schwarz iterative loop uses a different left hand side matrix. The subdomains do not update the left hand side matrix during Newton iterations. It uses the same matrix that was generated during the first Newton iteration. The advantage with that is that the $LU$ factors are computed (most time consuming step) only once and are repeatedly reused for all the inner and the outer iterative loops. This will increase the number of inner additive Schwarz iterations but nevertheless saves computational time.

# 5. PARDISO Solver

PARDISO package (part of Intel MKL library) is a high-performance, robust, memory efficient and easy to use software for solving large sparse symmetric and unsymmetric linear systems. The solver uses a combination of left- and right-looking Level-3 BLAS supernode techniques. PARDISO uses a Level-3 BLAS update and pipelining with a combination of left- and right-looking supernode techniques [14-16]. Unsymmetric permutation of rows is used to place large matrix entries on the diagonal. Complete block diagonal supernode pivoting allows dynamical interchanges of columns and rows during the factorization process. The level-3 BLAS efficiency is retained and an advanced two-level left–right looking scheduling scheme is used to achieve higher efficiency. The goal is to preprocess the coefficient matrix $A$ so as to obtain an equivalent system with a matrix that is better scaled and more diagonally dominant. This preprocessing reduces the need for partial pivoting, thereby speeding up the factorization process. METIS and Minimum degree ordering options are available within the solver.

## 6. Numerical Implementation

The solution of non-linear system of equations involves two iterative loops. One is the outer Newton's iterative loop (outer loop) and the other is the inner additive Schwarz domain decomposition loop (inner loop) to solve the linear system of equations (Eq. 9) simultaneously on different processors. Each subdomain problem in the inner loop is solved by a direct solver. Since the left hand matrix is not updated during the inner iterations, the resolution facility of direct solvers can be used to skip the factorization phase (i.e. only the solve phase is invoked). The inner iterations are continued till the error norm is within tolerance. In the outer iterative loop for Newton, the Jacobian is updated. Hence the factorization phase is invoked in each of the subdomain during the first inner iteration of all outer iterations. In the case of modified Newton, the Jacobian is not updated during the outer iterations. Consequently, the subdomains need not invoke the factorization phase during the first inner iteration loop of all the outer iterations. In summary, for the modified Newton algorithm, each of the subdomains perform *LU* factorization only once and all other subsequent calls to the solver invokes only the solve phase. In the case of Newton algorithm, each of the subdomains invokes the *LU* factorization during the first inner iteration of all the outer iterations. The performance of both Newton and modified Newton are examined in this paper. All the simulations are carried out on Ohio Supercomputing Cluster "Glenn". It is a cluster of AMD opteron multi-core, 2.6 GHZ, 8GB RAM machines.

## 7. Results and Discussion

Preliminary experiments have shown that METIS ordering performs better in terms of computational time and memory requirements. Hence METIS is used for ordering the entries of the matrix. Table 1 shows the comparison of CPU times and memory requirement for solving a 100x20x20 mesh on two processors for minimum degree (MD) and METIS ordering. In table 1, dof refers to the number of degrees of freedom (or the total number of equations). Results indicate that METIS performs much better than the minimum degree algorithm. Based on this observation, METIS ordering is used for all subsequent calculations.

Table 1 Performance of different orderings for PARDISO solver for a 100x20x20 mesh on 2 processors

| Ordering | #dof's | CPU time (sec) | Memory (GB) |
|---|---|---|---|
| MD | 44541 | 602 | 5.3 |
| METIS | 44541 | 178 | 1.6 |

Figure 2 shows the performance of NAS and NMAS algorithms on 2 processors for a 100x20x20 grid. The convergence rates of both the algorithms show Newton quadratic convergence. The residual norm is exactly similar for both the algorithms. The Newton based additive Schwarz algorithm takes 178 seconds and Modified Newton based additive Schwarz algorithm takes 87 seconds for full convergence. The time taken for the first Newton iteration is the same for both the algorithms but NMAS algorithms takes much lesser time for the subsequent iterations.

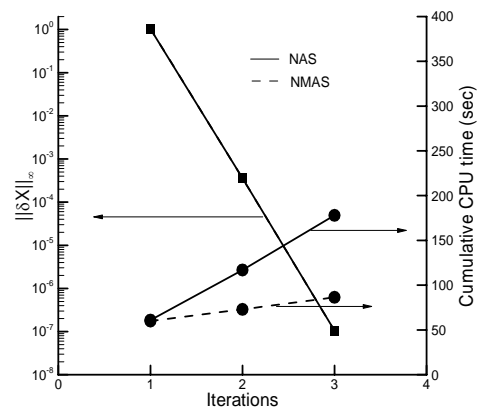

Figure 2: Comparison of CPU time and residual norm for NAS and NMAS algorithms for 100x20x20 grid on 2 processors.

Table 2 shows the performance of NAS and MNAS algorithms as a function of the number of processors. Excellent scalability is observed both for NAS and MNAS algorithms. For increase of processors from 2 to 12, the computational time reduced by a factor of 12.7 for NAS and a factor of 7.9 for MNAS algorithm. Figure 3 shows the pictorial representation of the scalability performance of the NAS and MNAS algorithms. Up to 8 processors, the scalability is good and beyond that the scalability decreases. For larger grids, it may be possible to obtain good scalability even beyond 8 processors. Note that the memory requirement for both the NAS and MNAS algorithms are exactly the same, since the size of the *LU* factors remain the same. It is observed that by increasing the number of processors from 2 to 12, the memory requirement on a single processor reduced by a factor of 11. Hence domain decomposition techniques are highly memory efficient. Unlike the use of a distributed parallel direct solver like MUMPS [6], the scalability of domain decomposition method is quite high. By using distributed computing using domain decomposition algorithm, relatively much larger problems can be solved. The memory limitation of direct solvers can be well addressed using domain decomposition technique by running the problem on a large number of processors.

Table 2: Comparison of the performance of NAS and MNAS for a 100x20x20 mesh

| # of processors | NAS CPU time (sec) | MNAS CPU time (sec) | max memory on one processor (MB) |
|---|---|---|---|
| 2 | 178 | 87 | 685 |
| 4 | 65.8 | 35.1 | 282 |
| 6 | 41 | 22 | 157 |
| 8 | 26 | 15 | 116 |
| 10 | 19 | 12 | 88 |
| 12 | 14 | 11 | 62 |

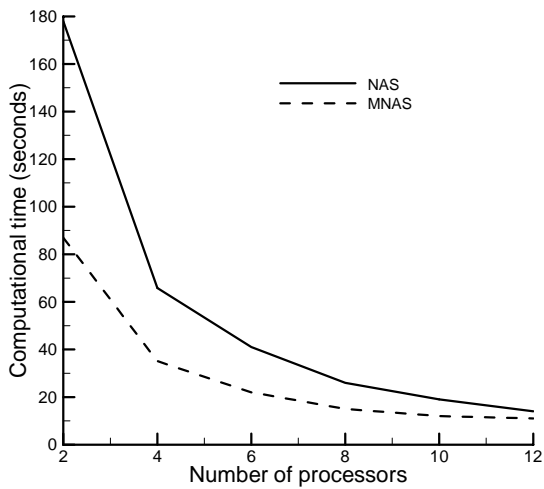

Figure 3: Scalability performance of NAS and MNAS on a 100x20x20 mesh

## 8. Conclusions

Distributed parallel computing of finite element Navier-Stokes code using additive Schwarz algorithm is demonstrated. An efficient sparse direct solver PARDISO is used for solving the subdomain problems. Two algorithms NAS and MNAS have been explored. It is observed that the MNAS algorithm can lead to significant savings in computational time. The additive Schwarz algorithm is found to scale well both in terms of computational time and memory requirement. It could of great value in large scale three dimensional computations.

## Acknowledgments

Author would like to thank the Ohio Supercomputing Centre for proving the computing facility.

**Mandhapati P. Raju** completed his MS (2002-2004) and Ph.D (2004-2006) in Mechanical Engineering department at Case Western Reserve University, Cleveland, OH. Later he worked as Postdoctoral fellow in Case Western Reserve University during 2006-2008. Later he worked at Caterpillar Champaign simulation centre as a CFD analyst. Currently he is working as a Post Doctoral fellow in General Motors Inc. His research interests are combustion, porous media flows, multifrontal solvers, fuel cell and hydrogen storage. This work was done during his presence in Case Western Reserve University. He has published 6 journal papers in reputed international journals.

**Siddhartha Kumar Khaitan** received the B.E. degree in electrical engineering from Birla Institute of technology, Mesra, India, in 2003 and the M.Tech. degree from the Indian Institute of Technology, Delhi, India, in 2005. He completed his PhD from Iowa State University in August 2008, and since then continuing as a Post Doctoral Research Associate at Iowa State University, Ames. His current research interests are power system dynamic simulation, cascading, numerical analysis, linear algebra, and parallel computing. Dr. Khaitan was the topper and Gold Medalist in his under-graduate studies and recipient of Research Excellence award for his PhD work.